\documentclass{article}
\usepackage[T1]{fontenc}
\usepackage[utf8]{inputenc}
\usepackage{ismir}
\usepackage{amsmath,amssymb,cite,url}
\usepackage{graphicx}
\usepackage{booktabs}
\usepackage{multirow}
\usepackage{makecell}
\usepackage{microtype}
\usepackage{tabularx}
\usepackage{color}
\usepackage{enumitem}
\setlist[itemize,1]{leftmargin=1.25em}

\title{SymphonyGen: 3D Hierarchical Orchestral Generation with Controllable Harmony Skeleton}

\multauthor
  {Xuzheng He$^1$ \hspace{1cm} Nan Nan$^2$ \hspace{1cm} Zhilin Wang$^3$ \hspace{1cm} Ziyue Kang$^2$}
  {
  {\bf Zhuoru Mo$^4$ \hspace{1cm} Ao Li$^2$ \hspace{1cm} Yu Pan$^1$}\\
  {\bf Xiaobing Li$^1$ \hspace{1cm} Feng Yu$^1$ \hspace{1cm} Xiaohong Guan$^{1,2}$}\\
  $^1$ Department of AI Music and Music Information Technology, Central Conservatory of Music\\
  $^2$ Frontier Institute of Science and Technology, and Interdisciplinary Research Center\\of Frontier Science and Technology, Xi'an Jiaotong University\\
  $^3$ University of Science and Technology of China, \hspace{0.1cm} $^4$ Shenzhen University\\
  {\tt\small 21sa026@mail.ccom.edu.cn, nan.nan@xjtu.edu.cn}
  }

\def\authorname{X. He, N. Nan, Z. Wang, Z. Kang, Z. Mo, A. Li, Y. Pan, X. Li, F. Yu, and X. Guan}

\usepackage[bookmarks=false,pdfauthor={\authorname},pdfsubject={\pdfsubject},hidelinks]{hyperref}

\begin{document}

\maketitle

\begin{abstract}
Generating symphonic music requires simultaneously managing high-level structural form and dense, multi-track orchestration, yet existing symbolic models often struggle with a ``complexity-control imbalance'' between scalability and steerability. We present SymphonyGen, a 3D hierarchical framework for contemporary orchestral generation, whose cascading decoders decompose the bar, track, and event axes, keeping decoding memory far below flat token streams and enabling conditioning at every structural level. A beat-quantized multi-pitch harmony skeleton, which may be user-written, analyzed, or model-generated, provides ``short-score'' conditioning, enabling outline control while producing orchestral textures. The model is refined with reinforcement learning against a cross-modal acoustic reward from CLaMP 3 audio embeddings, and a dissonance-averse sampling algorithm suppresses unintended tonal clashes during inference. Objective evaluations show that both post-training mechanisms reduce dissonance while maintaining independent melodic metrics, and in subjective tests SymphonyGen is rated above baseline systems in quality and preference, significantly so among general listeners.
\end{abstract}

\section{Introduction}
\begin{figure}
\centering
\includegraphics[alt={Overview diagram. User-written, analyzed, or model-generated sources produce a harmony skeleton, notated as beat-aligned chord columns on a grand staff. SymphonyGen expands the skeleton into a multi-instrument orchestral full score, steered by dissonance-averse sampling at inference; during training, GRPO refines the model with an acoustic reward from CLaMP 3 embeddings of reference soundtracks.},width=0.95\linewidth]{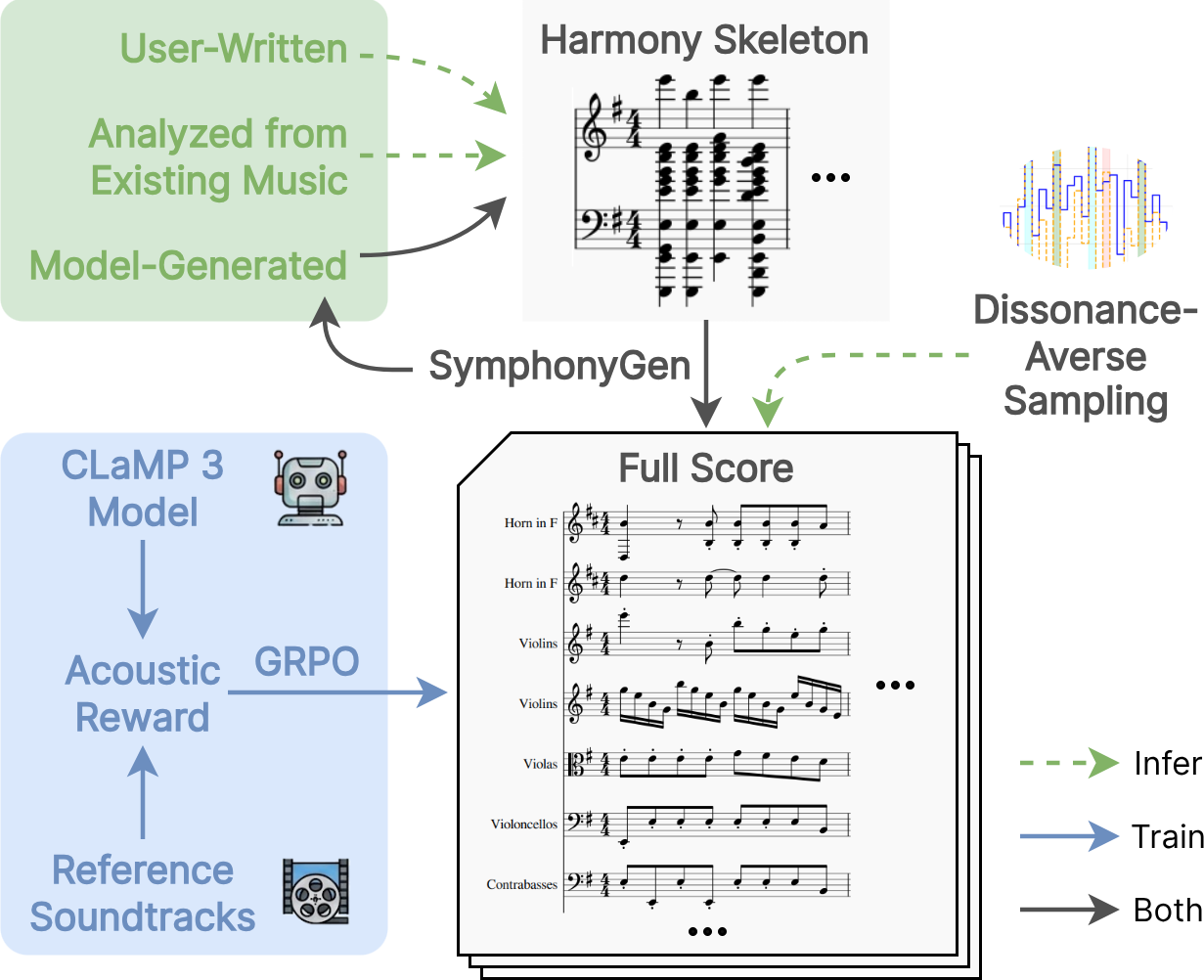}
\caption{Overview of the SymphonyGen system.}
\label{skeleton}
\end{figure}

The orchestra offers composers a rich palette of timbre and texture, and writing for it is a core discipline of Western composition practice~\cite{StudyOrchestration}.
Replicating it, however, remains formidable for symbolic AI: a model must capture dependencies across many voices while sustaining an evolving harmonic landscape over long horizons.
Considerable progress has been made through Transformer architectures~\cite{Transformer,MusicTransformer}, symbolic representations~\cite{REMI,CompoundWord}, and dedicated frameworks such as SymphonyNet~\cite{SymphonyNet} and NotaGen~\cite{NotaGen}.

While these models capture characteristic patterns of their target styles, they typically treat generation as a monolithic sequence-modeling problem that (i) scales poorly to dense orchestral textures, and (ii) bypasses the intermediate structural decisions that anchor the multi-stage workflows of professional production, such as drafting a short-score sketch~\cite{OnTheTrack}.
We frame the gap in controllable symphonic creation as a ``complexity-control imbalance''.

Existing controls roughly fall into three tiers: high-level cross-modal attributes such as style, scalar features, chord labels, or text~\cite{FIGARO,MuseMorphose,StylePriorArrange,Museformer,Text2midi,MetaScore}; content that prescribes verbatim note-level material, such as the melody in a lead sheet~\cite{AccoMontage,PopMAG,METEOR,Seq2SeqArrange}; and intermediate sketch stages~\cite{ThemeTransformer,CascadedDiffusion} like our work.
High-level controls can be phrase-level~\cite{AccoMontage,CascadedDiffusion}, bar-level~\cite{FIGARO,MuseMorphose}, or track-level~\cite{StylePriorArrange,METEOR}.
Prior work argues that the \emph{compositional hierarchy} of music is the key to handling these tiers and levels, in a cascaded diffusion model~\cite{CascadedDiffusion}.

A further obstacle is data quality: models trained on broad MIDI datasets inherit their noise and frequently produce dissonance (cf.\ the dissonance statistics of the dataset itself in \tabref{tab:objective}), at odds with the acoustic clarity we target.

To bridge these gaps, we present \textit{SymphonyGen}, a hierarchical framework controlled by a multi-pitch harmony skeleton for contemporary orchestral generation. While our architecture itself is style-agnostic, it is oriented toward contemporary cinematic scoring through the data, reward, and hyperparameters. Our contributions are as follows:
\begin{itemize}
\item \textbf{3D Hierarchical Architecture with Compressed Tokenization:} A cascading encoder-decoder system that processes the bar, track, and event axes separately, exposing conditioning points at each structural level and keeping decoding memory far below flat token streams; a compressed tokenization scheme reduces the event-axis padding inherent to the fixed 3D grid.
\item \textbf{Multi-Pitch Harmony Skeleton:} A novel ``short-score'' condition that provides a structural music outline, steering the composition through beat-quantized harmonic and melodic contours.
\item \textbf{Dissonance-Averse Sampling:} A logit-adjustment algorithm that references the harmony skeleton to suppress unintended clashes during inference, especially those involving non-harmonic tones.
\end{itemize}

\section{Related Work}
\subsection{Symbolic Orchestral Generation}
Generative models specifically targeting orchestral music are less common than those for general multi-track tasks. Early multi-track efforts like MuseGAN~\cite{MuseGAN} utilized GANs on piano-rolls before the field shifted toward Transformers over tokenized sequences. Pop Music Transformer~\cite{REMI} and Compound Word Transformer~\cite{CompoundWord} introduced fundamental tokenization schemes that remain influential, while MMM~\cite{MMM} and the Multitrack Music Transformer~\cite{MMT} explored track-structured token representations.

SymphonyNet~\cite{SymphonyNet} pioneered the task of symphonic generation and introduced a large-scale symphony dataset. Recently, NotaGen~\cite{NotaGen} proposed CLaMP-DPO, a symbolic-modality preference reward, to generate sheet music in a given classical style. While successful, they operate as single-stage generators without an intermediate, human-editable sketch stage.

\subsection{Music Orchestration and Accompaniment}
In the symphonic domain, early work trained conditional restricted Boltzmann machines~\cite{LOP} on a database of 196 paired piano--orchestra scores~\cite{LOPDatabase}.
In the pop domain, AccoMontage~\cite{AccoMontage,AccoMontage2} proposed a hybrid retrieval-based approach to piano accompaniment arrangement using phrase-labeled POP909 dataset~\cite{POP909}, while PopMAG~\cite{PopMAG} addressed end-to-end multi-track accompaniment, deriving the training data from a melody-track classifier.
MuseMorphose~\cite{MuseMorphose} introduced Transformer VAE for style transfer on pop piano, achieving bar-level attribute controls; METEOR~\cite{METEOR} extended this approach to the symphonic domain with melody control, supporting lead sheet orchestration.
Recent advances include long-term planning of per-track \emph{orchestral functions} (rhythmic-intensity patterns)~\cite{StylePriorArrange}, an intermediate stage of melody-reduced lead sheet generation~\cite{CascadedDiffusion}, and conditioning on a merged \emph{content sequence} of deduplicated notes~\cite{Seq2SeqArrange}.

While many conditions in prior work are numerical controls, our harmony skeleton is musical content that the user can sketch without full compositional expertise, e.g., by playing a few chords on a MIDI keyboard.

\subsection{Controllable and Structured Music Generation}
FIGARO~\cite{FIGARO} utilized bar-level interpretable features for controllable generation, while Theme Transformer~\cite{ThemeTransformer} focused on thematic conditioning. To model phrase structure, researchers proposed Museformer~\cite{Museformer} with structure-related attention and whole-song generation with hierarchical music language~\cite{CascadedDiffusion}.
Text2midi~\cite{Text2midi} and MetaScore~\cite{MetaScore} explored generating symbolic music from text.

While these methods offer diverse handles on music, our harmony skeleton adds a new way of outline-based control, playing a role in symphonic music comparable to a lead sheet in homophonic music, without presupposing the unique existence of a singable melody.

\section{Methodology}
\subsection{3D Hierarchical Architecture}\label{sec:arch}

\begin{figure*}
\includegraphics[alt={Four-phase architecture diagram. See text for detailed description.},width=\linewidth]{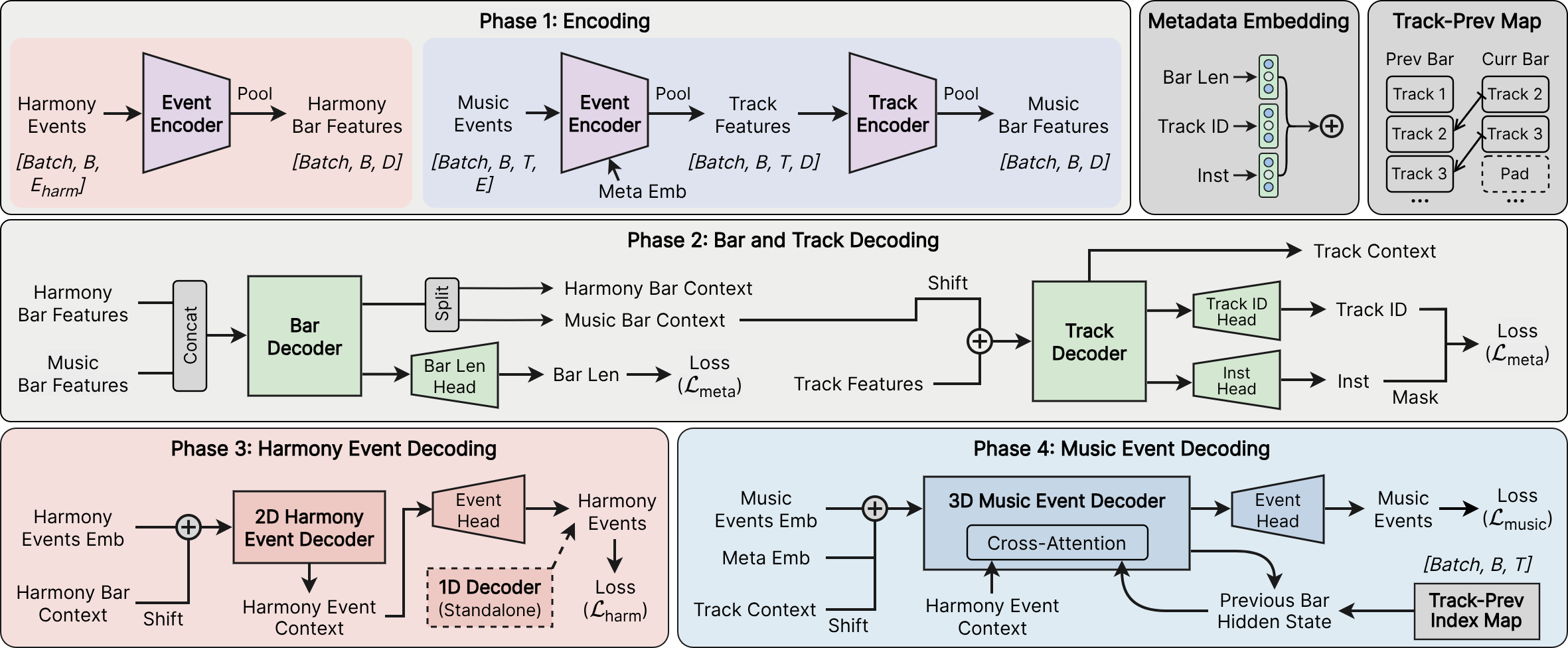}
\caption{Model architecture. The four phases are trained jointly with the total loss $\mathcal{L} = 0.05 \mathcal{L}_{\text{meta}} + 0.5 \mathcal{L}_{\text{harm}} + \mathcal{L}_{\text{music}}$.}
\label{architecture}
\end{figure*}

\begin{table}
  \small
  \setlength{\tabcolsep}{4pt}
  \centering
  \begin{tabular}{lll}
    \toprule
    Architecture & Attention Cost & Decode State \\
    \midrule
    \textbf{3D $[B,T,E]$} & $O(BTE^2 + BT^2 + B^2)$ & $O(B+T+E)^\dagger$ \\
    \textbf{2D $[BT, E]$} & $O(BTE^2 + B^2T^2)$ & $O(BT+E)$ \\
    \textbf{2D $[B, TE]$} & $O(BT^2E^2 + B^2)$ & $O(B+TE)$ \\
    \textbf{1D $[BTE]$} & $O(B^2T^2E^2)$ & $O(BTE)$ \\
    \bottomrule
  \end{tabular}
  \caption{Asymptotic comparison across architectures. $^\dagger$Plus an $O(TE)$ buffer for cross-attention (\secref{sec:twostream}).}
  \label{tab:complexity}
\end{table}

The backbone of our model consists of cascading Transformer encoders and decoders, each processing one musical dimension: $B$ bars, $T$ tracks, or $E$ note events.
We list the alternative dimensional configurations and compare with their asymptotic complexity in \tabref{tab:complexity}.
NotaGen~\cite{NotaGen} divides sheet notation into patches of 16 characters, akin to the 2D $[BT, E]$ model;
we focus on comparing with this 2D setup and the 1D model.

In a controlled efficiency benchmark at our scale ($B{=}T{=}E{=}32$),
our model decodes with approximately 17 times less cached state than a flat 1D layout, a memory gap that keeps growing with piece length.
Against 2D flattening, however,
wall-time is dominated by MLPs rather than attention. Because most bars have fewer than 32 tracks, a naive 3D model that pads to a fixed 32 tracks per bar takes approximately $2.5$ times the training time of a matched 2D model in our benchmark.
This training inefficiency can be addressed by packing the (bar, track) axes compactly in the event encoder/decoder and unpacking in other modules.
The 3D architecture can thus retain the efficiency of the 2D representation while providing more musically meaningful ways of allocating computation.

\subsection{Harmony Skeleton Analysis}\label{sec:skeleton}

We define the harmony skeleton as a beat-quantized ``short score'' of the pitches active at each beat (a quarter note), as shown in \figref{skeleton}.
It is extracted by rule-based analysis for training; at inference it could be user-written, analyzed from existing music, or model-generated.
The analysis consists of two steps:
\textbf{chord template matching}, which selects a template chord (any triad or seventh chord) by the highest cosine similarity of the pitch-class distribution at each beat;
and \textbf{extension identification}, which employs dynamic programming to include a maximal set of additional notes as chord extensions, under the constraint of no new minor or major second intervals being introduced.
The chordal notes and the extensions are stretched to beat boundaries to form the harmony skeleton, which is serialized with the same event vocabulary as in \secref{sec:tokenization}.

On the DCML harmony annotation corpus~\cite{DCMLCorpus}, the analysis achieves pitch-class precision/recall of 84.9\%/89.0\%, suggesting that the skeleton reasonably approximates the harmonic progression with detailed chord voicings.

\subsection{Encoding and Decoding}\label{sec:encdec}

As shown in \figref{architecture}, the forward pass of our model is grouped into four phases, described as follows:
\begin{itemize}
\item \textbf{Encoding (Phase 1).} Harmony events pass through an event encoder with pooling along the event axis to produce harmony bar features. Music events pass through the same encoder to produce track features, with metadata embeddings added (bar length, track ID, instrument).%
\footnote{An instrument may be shared by several tracks (e.g., Violin I/II).}
A track encoder with pooling processes the track features into music bar features.

\item \textbf{Bar and Track Decoding (Phase 2).} The harmony and music bar features are concatenated along the bar axis, and processed by a bar decoder, then split back into harmony bar context and music bar context.
By reviewing the harmony bar features from beginning to end, the bar decoder can form a bar-level plan that guides lower hierarchies.
Specifically, after being right-shifted from the previous bar, the music bar context is propagated down to the track features.   A track decoder then processes these features into track context.

\item \textbf{Harmony and Music Event Decoding (Phase 3--4).} Analogous to NotaGen's character-level decoder, harmony and music events are decoded locally along the event axis, guided by the (higher-level) harmony bar context or the track context. A 2D harmony event decoder produces harmony event context, which is attended by a 3D music event decoder. A standalone 1D decoder is used to decode harmony events at inference.
\end{itemize}

The model is trained end-to-end with three loss components: $\mathcal{L}_{\text{meta}}$ for metadata predictions, $\mathcal{L}_{\text{harm}}$ for harmony event prediction, and $\mathcal{L}_{\text{music}}$ for music event prediction.
The harmony skeleton condition is injected at two levels: harmony bar features prepended to the music bar features, and harmony event context attended by the music event decoder.
We refer to the inference procedure in our codebase.\footnote{\url{https://github.com/symphonygen/symphonygen}\label{fn:code}}

\subsection{Two-Stream Cross-Attention}\label{sec:twostream}

The music event decoder employs a two-stream cross-attention mechanism: odd layers attend to the harmony event context of the current bar, while even layers attend to the hidden states retrieved from the previous bar for the same track ID.
Due to varying active tracks across bars, we use an auxiliary index map to link each track to its previous-bar counterpart in parallel during training (\figref{architecture}).

In a controlled comparison, the two streams reduced validation loss by about 30\% compared to using either stream alone, suggesting that both contribute to mitigating the information bottleneck of the hierarchical model.

\subsection{Tokenization and Metadata Handling}\label{sec:tokenization}

\begin{figure}
\centering
\includegraphics[alt={Two note timelines with token lists before and after compression. Note Grouping: C4 and D4 sounding at the same position with equal duration share a single position and duration token, and the bar-initial position token is omitted. Legato Fusion: when a note begins exactly where the previous one ends, a single Legato token replaces that note's duration token and the following position token.},width=\linewidth]{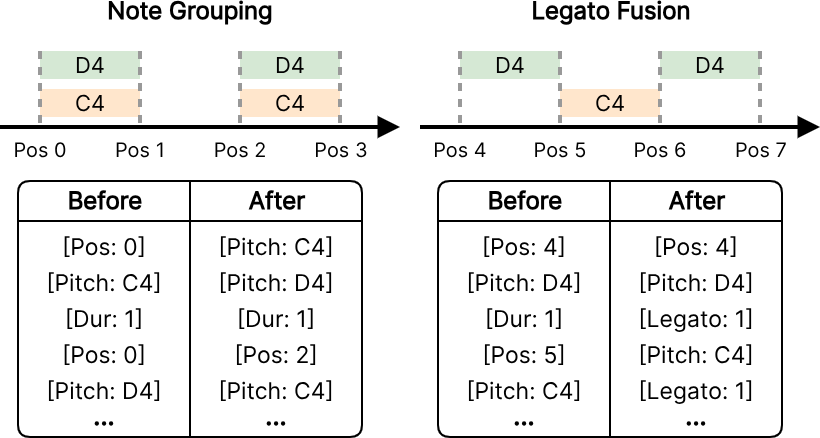}
\caption{Tokenization compression scheme.}
\label{tokenization}
\end{figure}

To balance padding and truncation of the event axis, we propose a compressed REMI~\cite{REMI} tokenization scheme. It utilizes five types of event tokens: Pitch, Position, Duration, End-Of-Sequence, and a novel \textit{Legato}, on a quantized 32nd-note grid. The compression involves two main techniques (\figref{tokenization}):
\textbf{note grouping}, which shares the position token for notes at the same position and the duration token for notes with also the same duration;
and \textbf{legato fusion}, which compresses tokens where a note starts immediately after the previous one ends.
A \texttt{[Legato:\,d]} token at position $p$ replaces \texttt{[Dur:\,d]} and \texttt{[Pos:\,p+d]}, resembling a byte-pair encoding (BPE) merge rule but fixed and context-sensitive rather than learned and context-free.
We also omit the downbeat position token \texttt{[Pos:\,0]}, treating it as the implicit default position for each bar.

Under our $B{=}T{=}E{=}32$ setting, with $E_{\text{harm}}=64$ harmony events per bar, the data fit within their respective limits in 97\% of music (bar, track) pairs and 98\% of harmony bars.
The vanilla tokenization requires doubling $E$ to cover cases such as a continuous run of sixteenth notes.

SymphonyGen predicts the metadata using separate heads of the bar decoder and the track decoder.
The track ID head is responsible for ending a bar with a special label, akin to the End-of-Sequence special token.
The instrument head is masked for existing tracks whose instrument is already assigned at the track's first occurrence.

\subsection{Reinforcement Learning via GRPO}\label{sec:grpo}

After pretraining, the model is refined with reinforcement learning via Group Relative Policy Optimization (GRPO)~\cite{GRPO}.
In each epoch, $K$ harmony skeletons are sampled. For each skeleton, a group of $G$ musical pieces are sampled from the model.
Advantages of each piece are estimated by normalizing rewards within the group, effectively comparing different orchestrations of the same harmony skeleton.
Details on the GRPO algorithm and its loss function can be found in the original paper~\cite{GRPO}; here we focus on describing the reward function.

We utilize the audio encoder of CLaMP 3~\cite{CLaMP3}, a three-way cross-modal alignment model between sheet music, audio, and text.
Each musical piece is first synthesized by MuseScore~3, then embedded by CLaMP 3 with average pooling over 5-second windows.
The cosine similarity score against the centroid embedding of a reference audio set serves as the reward.

This setup is a variant of NotaGen's CLaMP-DPO, which uses CLaMP 2 sheet music embeddings.
In their work, the entire reference set is partitioned into 112 labels (Period, Composer, Instrumentation) for style-conditioned generation, whereas we have so far experimented with only a single style derived from a few reference tracks.

\subsection{Dissonance-Averse Sampling}

We implement inference-time dissonance-averse sampling relative to a reference harmony skeleton. Since the harmony skeleton discerns harmonic ($H$) and non-harmonic ($N$) tones at any time step $t$, we can distinguish between different types of dissonances.

First, we define a symmetric dissonance matrix $W$, where $W_{ij}\in [0, 1]$ represents the degree of clash between pitch $i$ and pitch $j$.
Then, we compute the following score:
\begin{equation}\label{dissonance}
D_{\text{total}} = \lambda_{hn} \sum_{t} H_t^\top W N_t + \lambda_{nn} \sum_{t} \frac{1}{2} N_t^\top W N_t
\end{equation}
where $H_t$ and $N_t$ are the normalized occupancy vectors of active harmonic and non-harmonic tones, respectively.

The values in $W$ are guided by interval sensory distance as described in the Plomp-Levelt curve~\cite{PlompLevelt}; we set these weights manually, where (for example) minor seconds and tritones receive weights of $1.0$ and $0.95$, while perfect fifths receive $0.1$.
To improve acoustic clarity, $W$ can be further modulated by a register-dependent decay function to respect the Low Interval Limit---a principle in orchestration that necessitates wider intervals in lower registers.

During music event decoding, we subtract the logits of pitch tokens by the amount of dissonance each pitch would introduce at that time. The pitch logits are then re-normalized, preserving the total probability mass. Dissonance aversion reflects a stylistic prior rather than a universal musical criterion.

\section{Experimental Setup}
\subsection{Dataset}
We utilize the \textbf{SymphonyNet Dataset}~\cite{SymphonyNet}, comprising 728 classical and 45,632 contemporary MIDI files.
The MIDI files are split into 90\% training and 10\% validation sets.

\subsection{Training and Inference Details}\label{sec:details}
SymphonyGen is implemented with 33 layers (512 hidden size, 124M parameters), consisting of 8 harmony event decoder layers, 9 music event decoder layers, and 4 layers for each of the remaining encoders and decoders.
Pre-training was conducted on four NVIDIA H800 GPUs for one day using the AdamW optimizer~\cite{AdamW} ($lr = 1 \times 10^{-4}$) with cosine annealing.
GRPO was conducted on a single GPU until the reward reached a plateau ($lr = 4 \times 10^{-5}$), using $K=16$ and $G=32$.
The CLaMP 3 reference set consists of five game cinematic music tracks.

In both the GRPO rollout and final inference, harmony skeletons are generated via a standalone 12-layer 1D decoder (768 hidden size, 87M parameters).
We employ filters rejecting skeletons with low density ($<3$ notes/beat), excessive repetition ($>25\%$), or unusual log probability.
Skeletons that pass these filters are pruned into pure template chords, since the current model is not relied upon for generating extensions. This step is unnecessary for user-written or analyzed skeletons, and future work could refine it by incorporating voice-leading constraints.

We set sampling parameters $(\lambda_{hn},\lambda_{nn})$ to $(1,10)$, top-$p$ to $0.99$ and temperature to $1.0$.
Pitch logits out of the instrument ranges based on statistics of the dataset are masked in music event decoding.
The tempo is a constant $120$ BPM.

\subsection{Objective Metrics}\label{sec:metrics}
We report the following objective metrics for our model:
\begin{itemize}
\item \textbf{Monitoring Metrics.} \textit{CLaMP} is the similarity score used as the RL reward (\secref{sec:grpo}); \textit{Dissonance} ($D_{hn}, D_{nn}$) is the mean of $H_t^\top W N_t$ and $N_t^\top W N_t / 2$ across time steps (\eqnref{dissonance}); and \textit{Track Density} (Trk) is the mean number of active tracks.

\item \textbf{Harmony Precision \& Recall (Prc, Rec)} is the pitch overlap between the input harmony skeleton and the skeleton re-analyzed from the generated output.

\item \textbf{Melodic Movement (Mov)} measures the frequency of rhythmic shifts based on changes in the predominant durations of ``skyline'' tracks between adjacent bars.

\item \textbf{Melodic Ornament (Orn)} measures the presence of three passing tones (eighth or sixteenth notes) that precede a sustained note in the same track.
\end{itemize}

The monitoring metrics are either directly optimized or strongly affected (e.g., Trk by the skeleton density filter); our discussion focuses on how controlling one metric affects another.
The melodic metrics (Mov, Orn) are crafted by us to assess melodic and rhythmic vitality, complementing the harmonic metrics.
They guard against the model discarding interesting melodic content, under the assumption that higher scores within certain ranges indicate closer alignment with our target style.
Detailed implementations are available in our codebase.

\subsection{Subjective Test Design}\label{sec:subjective}
We conduct two rounds of subjective listening tests. The first test focuses on composition, and the second on re-orchestration.
For the first test, we generate one sample for each of 24 model-generated skeletons, with additional skeleton filters requiring at least four $V/V^{(7)} \to I/i$ cadences in 32 bars and a major or minor opening chord.
These filters emulate a user curating candidate skeletons for tonal closure.
For the second test, we generate one sample for each of 10 randomly selected skeletons from the validation set.
The baselines are as follows:
\begin{itemize}
\item \textbf{SymphonyNet (42M)~\cite{SymphonyNet}.} We include seven official demo samples in the orchestral style (30 seconds to two minutes), either conditioned on a per-measure chord progression or generated from scratch.

\item \textbf{NotaGen (516M)~\cite{NotaGen}.} We compare with its latest version hosted online, NotaGen-X. Since it does not support contemporary styles, we set the label to (Romantic, Brahms/Dvořák, Orchestral), with Romantic being the closest available style to contemporary, and Brahms/Dvořák more frequently producing orchestral textures. We sample 29 windows of 32 bars that contain more than 6 tracks, with a stride of 16 bars.

\item \textbf{METEOR (67M)~\cite{METEOR}.} We include three official demo samples (8 bars each) in the orchestral style, two conditioned on dataset excerpts and one on a lead sheet.
\end{itemize}

We resort to demo samples for SymphonyNet and METEOR because their implementations require input MIDIs (a primer for continuation or a latent code for re-orchestration) unavailable in our conditioning framework. The comparisons are therefore indicative benchmarks rather than fully matched experiments.
We also compare with four orchestral excerpts from the dataset, selected from sampled files that underwent the same 32-bar window filtering as NotaGen.
Participants are asked to rate at least three anonymized excerpts on a 5-point Likert scale across four dimensions: \textbf{overall quality}, \textbf{musical coherence}, \textbf{orchestration richness}, and \textbf{preference}.
All test excerpts are available on our demo page.

\section{Results and Discussion}
\begin{table*}[t!]
\small
\setlength{\tabcolsep}{6pt}
\centering
\begin{tabular}{l|c|cccc|cccc}
\toprule
 & $(\lambda_{hn}, \lambda_{nn})$ & CLaMP & $D_{hn}$ & $D_{nn}$ & Trk & Prc & Rec & Mov & Orn \\
\midrule
Dataset & - & .473$_{\pm.096}$ & .695$_{\pm.507}$ & .064$_{\pm.089}$ & 10.75$_{\pm5.70}$ & 1.00$_{\pm0}$ & 1.00$_{\pm0}$ & .198$_{\pm.145}$ & .108$_{\pm.109}$ \\
\midrule
NotaGen & - & .387$_{\pm.079}$ & .381$_{\pm.245}$ & .029$_{\pm.031}$ & \phantom{1}5.52$_{\pm2.06}$ & - & - & .225$_{\pm.153}$ & .058$_{\pm.082}$ \\
\midrule
Ours & (0, 0) & .589$_{\pm.065}$ & .777$_{\pm.443}$ & .074$_{\pm.072}$ & 15.35$_{\pm6.89}$ & .940$_{\pm.037}$ & .818$_{\pm.063}$ & .257$_{\pm.130}$ & .099$_{\pm.102}$ \\
\midrule
\multirow{4}{*}{\makecell[l]{Ours\\(with RL)}} & (0, 0) & .726$_{\pm.044}$ & .402$_{\pm.309}$ & .031$_{\pm.066}$ & \phantom{1}6.04$_{\pm1.66}$ & .948$_{\pm.043}$ & .706$_{\pm.084}$ & .288$_{\pm.137}$ & .102$_{\pm.089}$ \\
 & (1, 2) & .724$_{\pm.041}$ & .248$_{\pm.187}$ & .014$_{\pm.020}$ & \phantom{1}6.15$_{\pm1.78}$ & .959$_{\pm.043}$ & .712$_{\pm.093}$ & .294$_{\pm.141}$ & .097$_{\pm.088}$ \\
 & (1, 10) & .726$_{\pm.039}$ & .236$_{\pm.161}$ & .009$_{\pm.016}$ & \phantom{1}5.94$_{\pm1.52}$ & \textbf{.959}$_{\pm.041}$ & \textbf{.712}$_{\pm.082}$ & \textbf{.300}$_{\pm.132}$ & \textbf{.113}$_{\pm.093}$ \\
 & (5, 20) & .720$_{\pm.041}$ & .159$_{\pm.112}$ & .008$_{\pm.014}$ & \phantom{1}6.20$_{\pm2.03}$ & .948$_{\pm.053}$ & .693$_{\pm.093}$ & .288$_{\pm.136}$ & .098$_{\pm.096}$ \\
\bottomrule
\end{tabular}\caption{Objective results over 200 samples for each system. SymphonyNet and METEOR are not reported, as their codebases require MIDI inputs; the metrics would thus reflect partially the properties of the inputs rather than the capabilities of the model. For NotaGen, dissonance is computed relative to the analyzed harmony skeleton.}
\label{tab:objective}
\end{table*}

\begin{table*}[t!]
\small
\centering
\begin{tabular}{l|cccc|cccc}
\toprule
& Quality & Coherence & Richness & Preference & Quality & Coherence & Richness & Preference \\
\midrule
Dataset & 3.73$_{\pm0.91}$ & 3.63$_{\pm0.85}$ & 3.70$_{\pm0.99}$ & 3.30$_{\pm1.12}$ & 3.53$_{\pm1.12}$ & 3.82$_{\pm0.81}$ & 3.59$_{\pm1.00}$ & 3.24$_{\pm0.97}$ \\
\midrule
SymphonyNet & \underline{3.34}$_{\pm0.97}$ & 3.09$_{\pm1.18}$ & \underline{3.42}$_{\pm1.09}$ & \underline{2.82}$_{\pm1.10}$ & \underline{3.27}$_{\pm1.03}$ & 3.13$_{\pm0.83}$ & 3.20$_{\pm0.94}$ & \underline{3.00}$_{\pm1.11}$ \\
NotaGen & 3.23$_{\pm1.06}$ & \underline{3.43}$_{\pm1.20}$ & 3.11$_{\pm1.11}$ & 2.80$_{\pm1.28}$ & 3.23$_{\pm1.01}$ & \textbf{3.46}$_{\pm1.20}$ & \underline{3.31}$_{\pm1.03}$ & 2.85$_{\pm0.99}$ \\
Ours & \textbf{3.84}$_{\pm0.91}$ & \textbf{3.95}$_{\pm0.84}$ & \textbf{3.50}$_{\pm0.94}$ & \textbf{3.55}$_{\pm0.97}$ & \textbf{3.43}$_{\pm0.86}$ & \underline{3.43}$_{\pm0.90}$ & \textbf{3.37}$_{\pm0.72}$ & \textbf{3.10}$_{\pm0.80}$ \\
\midrule
$p$-value & 0.015 & 0.013 & 0.723 & 0.001 & 0.569 & 0.932 & 0.830 & 0.736 \\
\bottomrule
\end{tabular}
\caption{Subjective results for composition, split by general listeners (left) and educated listeners (right). The $p$-value of a two-sided Student's $t$-test between the best (bold) and second-best (underlined) models in each column is reported.}
\label{tab:generation}
\end{table*}

\begin{table}[t!]
\small
\centering
\setlength{\tabcolsep}{4pt}
\begin{tabular}{lcccc}
\toprule
& Quality & Coherence & Richness & Preference \\
\midrule
Dataset & 3.32$_{\pm1.01}$ & 3.56$_{\pm0.89}$ & 3.50$_{\pm1.08}$ & 3.06$_{\pm1.04}$ \\
\midrule
METEOR & 2.67$_{\pm1.02}$ & 3.15$_{\pm1.16}$ & 2.62$_{\pm0.89}$ & 2.45$_{\pm1.12}$ \\
Ours & \textbf{3.09}$_{\pm1.14}$ & \textbf{3.38}$_{\pm1.16}$ & \textbf{3.00}$_{\pm1.15}$ & \textbf{2.74}$_{\pm1.19}$ \\
\midrule
$p$-value & 0.116 & 0.405 & 0.131 & 0.324 \\
\bottomrule
\end{tabular}
\caption{Subjective results for arrangement.}
\label{tab:arrangement}
\end{table}

\subsection{Objective Evaluation}\label{sec:objective}

As shown in \tabref{tab:objective}, reinforcement learning (RL) reduces dissonance by nearly half, with a slight Mov increase and stable Orn.
This indicates that our acoustic reward improves harmonic clarity without diminishing melodic vitality.
The main side effect is a substantial drop in Trk, which we later counteract with a composite reward (\secref{sec:round2}).
Dissonance-averse sampling further suppresses dissonance while preserving CLaMP similarity.

As $\lambda_{hn}$ and $\lambda_{nn}$ increase, all reported harmonic and melodic metrics reach their peak around $(\lambda_{hn}, \lambda_{nn}) = (1, 10)$.
Beyond this point (e.g., at $(5, 20)$), the model outputs become distorted and mechanical in character.
We therefore adopt $(1, 10)$ as the setting for subjective tests.

\subsection{Subjective Evaluation}

\subsubsection{Round 1: Orchestral Composition Task}
We collected 237 evaluations from 32 participants (10 of them had a musical background, referred to as ``educated listeners''). As shown in \tabref{tab:generation}, SymphonyGen receives the highest average rating in seven of the eight columns.

\textbf{General Listeners:}
Our model achieves a significant lead in quality, coherence, and preference within this group.
Notably, it surpasses the ratings of the selected dataset excerpts in all categories except richness.
We attribute this improvement over the dataset examples to a combination of our design choices, including skeleton filtering, dissonance-aware sampling, and reinforcement learning, though their individual contributions are not isolated in this study.

\textbf{Educated Listeners:}
While SymphonyGen remains the highest-rated model on average, the performance gap narrows among educated listeners and no dimension reaches significance.
One possible explanation is that educated listeners are more tolerant of harmonic tension and attend more to structural intent than to surface content.

\subsubsection{Round 2: Orchestral Arrangement Task}\label{sec:round2}
To address the track density drop (\secref{sec:objective}) and given our slight lead in orchestral richness in the first round, we incorporated Trk into the reward (e.g., $\text{CLaMP 3} + 0.2 \times \tanh(\text{Trk} / 4)$), which increased it to $7.78\pm2.08$.
We also activated the register-dependent decay in $W$ to maintain bass clarity as track density rose.

We collected 102 evaluations from 17 participants (all educated listeners).
As shown in \tabref{tab:arrangement}, SymphonyGen is rated above the METEOR baseline in all dimensions, with the largest margins in quality and richness, although no single dimension reaches significance at this sample size.
This suggests that, when augmented with a track density reward, our model achieves orchestration comparable to a dedicated arrangement model.

\subsubsection{Qualitative Feedback}

We collected the participants' free-text comments to better understand our model's performance and limitations.

\textbf{Strengths:} Participants frequently described the style of our model using phrases such as ``a strong sense of fate'', ``storytelling'', or as having evocative imagery, consistent with our RL steering toward the cinematic reference set.

\textbf{Weaknesses:} Listeners still occasionally found that the ``harmony was strange'' or contained ``many mistakes'', with others describing specific segments as ``noisy'' or ``messy''.
These issues may stem from errors in harmony skeleton generation.
In the Round 2 test, some participants noted that the orchestration could feel ``too full''.
This highlights a challenge in manual reward shaping.

\section{Conclusion}\label{sec:conclusion}
We introduce SymphonyGen, a 3D hierarchical framework for orchestral generation.
Decomposing the bar, track, and event axes exposes conditioning at every structural level while keeping decoding memory low.
The multi-pitch harmony skeleton steers the composition and enables re-orchestration.
Refined by reinforcement learning and dissonance-averse sampling, SymphonyGen produces orchestral scores oriented toward a cinematic reference style, and was rated above baseline systems in our listening tests, significantly so among general listeners.

Future work includes skeleton generation with voice-leading constraints; multi-style reference audio sets; a matched 2D-backbone comparison; cross-attention ablations; measuring diversity under a shared skeleton; and evaluating SymphonyGen as a composition-support tool, e.g., with the Creativity Support Index~\cite{CSI}.

\section{Acknowledgments}
This work was supported by the National Natural Science Foundation of China under Grant T2341003.
We thank the anonymous reviewers for their constructive feedback.

\section{Ethics Statement}

Our training data comprises publicly available symbolic music datasets and audio references for reward modeling. SymphonyGen is intended as a collaborative aid for composers, not as an autonomous replacement. We expect human composers to derive the greatest benefit from its controllability. Users should also be mindful that generative outputs may inadvertently resemble existing works, and should not assume all outputs are fully original without verification.

\bibliography{ISMIRtemplate}

\end{document}